\documentclass[aps,twocolumn]{revtex4}
\usepackage{amsmath,amssymb}

\usepackage{graphicx}% Include figure files
\usepackage{dcolumn}% Align table columns on decimal point
\usepackage{bm}% bold math
\usepackage{hyperref} % Hiperlinks

\begin{document}

\title{Direct measurements of band gap grading in polycrystalline CIGS solar cells}

\author{M. P. Heinrich$^{1,2}$}
\author{Z-H. Zhang$^{1,2}$}
\author{Y. Zhang$^1$}
\author{O. Kiowski$^2$}
\author{M. Powalla$^2$}
\author{U. Lemmer$^1$}
\author{A. Slobodskyy$^{1,2}$}
\affiliation{$^1$Light Technology Institute (LTI), Karlsruhe Institute of Technology (KIT), Kaiserstr. 12, 76131 Karlsruhe, Germany\\
 $^2$Zentrum f\"{u}r Sonnenenergie- und Wasserstoff-Forschung Baden-W\"{u}rttemberg, Industriestr. 6, 70565 Stuttgart, Germany}

\date{\today}

\begin{abstract}
We present direct measurements of depth-resolved band gap variations of CuIn$_{1-x}$Ga$_x$Se$_2$ thin-film solar cell absorbers. A new measurement technique combining parallel measurements of local thin-film interference and spectral photoluminescence was developed for this purpose. We find sample-dependent correlation parameters between measured band gap depth and composition profiles, and emphasize the importance of direct measurements. These results bring a quantitative insight into the electronic properties of the solar cells and open a new way to analyze parameters that determine the efficiency of solar cells.
\end{abstract}


\maketitle

Variation of the band gap within the depth profile of solar cells is known as band gap grading and is a long-discussed approach for efficiency improvement \cite{Contreras1996}. Despite this, relatively few direct measurements were done to compare a real band gap profile with the one expected from the material composition. Moreover, no such measurements performed on high-efficiency thin-film solar cells have been reported.

Copper indium gallium diselenide (CIGS) is one of the most promising materials for application in highly efficient thin-film solar cells \cite{Contreras1999}.  A high light absorption coefficient and a tunable band gap are the two most remarkable characteristics of the CIGS absorber layers. The band gap of the CIGS absorber increases for higher Ga/(Ga+In) (GGI) ratio. Record efficiencies above 20\% have been achieved~\cite{Repins2008} by utilizing a graded band gap~\cite{Contreras1999}. Figure~\ref{fig-rem}~(a) shows a scanning electron microscope (SEM) image of a CIGS layer used in our experiments.

The mass content of Ga and In within the CIGS layer can be determined using secondary neutral mass spectroscopy (SNMS) \cite{Dullweber2001}, Raman line scan on a cross section \cite{Witte2006, Izquierdo-Roca2009} or  at different growth times \cite{Fontane2009}. However these methods suffer from several inaccuracies.  One of the most important is the fact that calculations of band gap energies with the GGI ratio $x$ rely on an empirically derived equation~\cite{Dimmler1987}:

\begin{figure}[t]
\centerline{\includegraphics[width=3.375in]{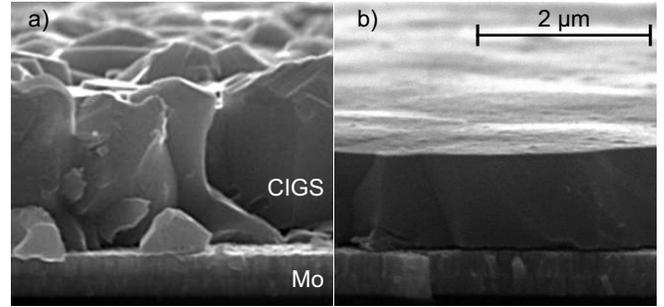}}
\caption{\label{fig-rem}SEM images of a CIGS layer on top of a 500~nm Mo layer.  (a)~Before etching. (b)~After etching with aqua regia. The layer thicknesses are 2 $\mu$m and 1 $\mu$m respectively.}
\end{figure}

\begin{equation}
E_g(eV)=1.02+0.66x-bx(1-x) 
\label{eq_bandgap}
\end{equation}

The bowing factor $b$ in this equation, which has to be known to get quantitative results, varies strongly in the literature, i.e. the band gap profile cannot be calculated reliably  from composition measurements.

Spectral cathodoluminescence line scans near the surface (depth of less than 400~nm) of CIGS layers were previously performed to detect a surface layer with a higher band gap \cite{Romero2003}, but no measurements of the Ga grading profile have been done so far.

The investigated CIGS samples were fabricated by an in-line~\cite{Powalla2009} (sample~A) and by a stationary (sample~B) co-evaporation multi-stage process. The structure of the samples is similar to the one described for the record efficiencies CIGS solar cells~\cite{Voorwinden2003}. The absorber layer thickness was 2.2~$\mu m$ in sample~A and 3.1~$\mu m$ in sample~B. Mean values for the GGI ratio are close to 0.3.

The front ZnO:Al contact and the CdS buffer layer of the solar cells are removed by a selective etching process with hydrochloric acid; a SEM image of the resulting structure is shown in Fig.~\ref{fig-rem}~(a). Etching of the CIGS absorber layer is possible with nitric acid and has been previously demonstrated \cite{Batchelor2004}. The most homogeneously etched layers are obtained in our experiments using a mixture of hydrochloric and nitric acid, which forms aqua regia. The molar mixing ratio 1:3 of HCl to HNO$_3$ resulted in etching rates of roughly 10~nm/sec, 3 minutes after preparation of the acid.  Step profiles of different layer thicknesses were processed. Figure~\ref{fig-rem}~(b) shows a SEM cross section of an etched absorber layer: in addition to a reduced layer thickness, a significant planarization of the surface roughness is achieved.

An optical experimental setup for combined measurements of thin-film interference and PL spectroscopy is realized using a confocal scanning microscope. A 532~nm laser is used as a PL excitation source. The absorption length of the excitation light in the CIGS layer with $k=0.55 $\cite{Paulson2003} is about 80~nm. The laser and a broadband light source were directed into the microscope objective and focused onto the sample. The measurements are done with a thermoelectrically cooled near-infrared array photodetector spectrometer that has a spectral sensitivity in the wavelength range from 900~nm to 1700~nm.

\begin{figure}[t]
\centerline{\includegraphics[width=3.375in]{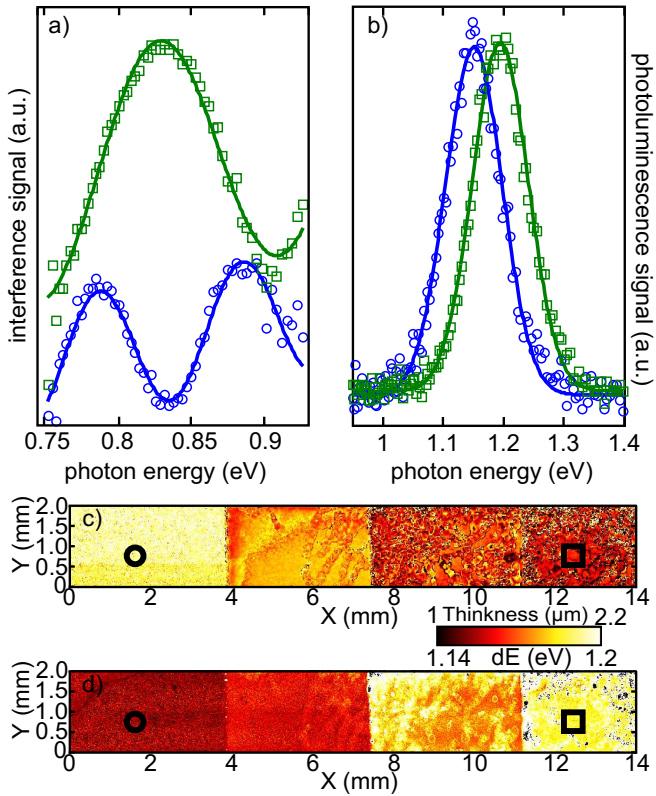}}
\caption{\label{spatial} (a) and (b) are interference spectra room temperature PL at two different locations ((Loc.~1: $\bigcirc$) and ((Loc.~2: $\square$). (c) and (d) are mappings of the absorber layer thickness the optical band gap at the same locations on CIGS sample~A with a gradually etched depth profile and a spatial resolution of 10~$\mu$m.}
\end{figure}

Determination of thin-film thicknesses using interference spectroscopy is a highly accurate, fast and non-destructive characterization method. The spectral distribution of the interference intensity $I(\hbar\omega)$ is approximated by the equation:

\begin{equation}
I(\hbar\omega)=a\cdot cos(2\pi \cdot2dn\frac{\omega}{2\pi c_0})+b+c\cdot\hbar\omega,
\label{eq_interference}
\end{equation}

where $d$ is the film thickness, $n$ is the refractive index and $c_0$ is the speed of light in vacuum. The parameters $a$, $b$ and $c$ depend on the reflection, absorption and scattering coefficients of the interfaces and materials and are empirically determined for each individual measurement.

In Fig.~\ref{spatial}~(a) results of the thin-film interference signal at different locations on the sample~A are shown. The spectra were fitted using Eq.~\ref{eq_interference} and shown by continuous lines in the figure. A constant refractive index of $n=3$ was used for the calculations \cite{Paulson2003}. Reflection coefficients for the interfaces between CIGS/air and CIGS/Mo are $0.23$ and $0.73$, respectively. The layer thicknesses determined by interference spectroscopy correspond very well to the values obtained from a Dektak profilometer.

We find the local optical band gap of the material from a local PL spectrum peak position by a Gaussian fit as shown in Fig.~\ref{spatial}~(b).

In Fig.~\ref{spatial}~(c) a two-dimensional absorber layer thickness profile of sample~A is shown in a color plot. The spatial resolution of the scan is 10~$\mu$m. Etching inhomogeneities increase with decreasing layer thickness and result from internal reaction in the aqua regia.

\begin{figure}[t]
\centerline{\includegraphics[width=3.375in]{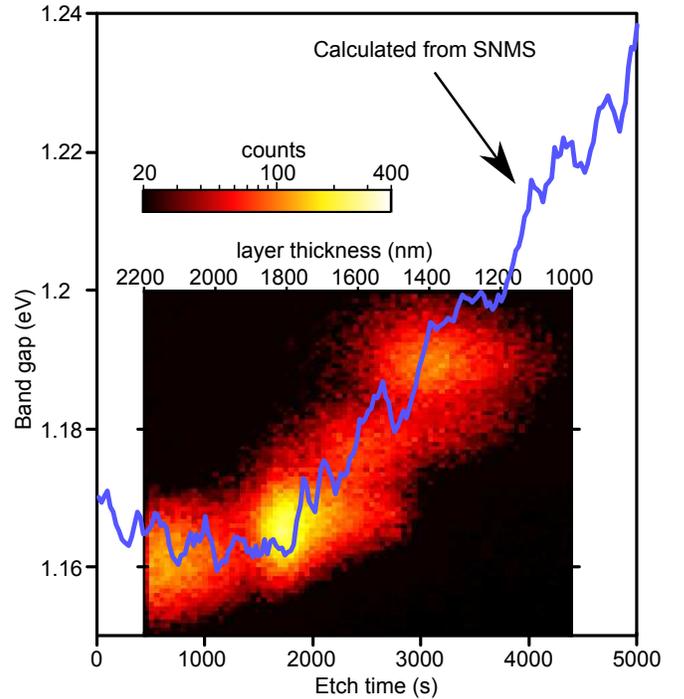}}
\caption{\label{histogramma}Grading profile of the band gap versus the absorber depth, obtained from the 2D histogram of the spatially resolved mapping data (the color plot) and calculated from the SNMS sputtering time dependence of the GGI ratio (solid line). The sample~A - 280,000 spectral measurements.}
\end{figure}

In Fig.~\ref{spatial}~(d) a two-dimensional scan of the band gap energies is shown on the same area as the layer thickness measurements in Fig.~\ref{spatial}~(a). It is possible to see that band gap is increasing with decreasing layer thickness.

We count the occurrence of points with values of the local band gap and the local layer thickness within a specific range over the sample area, in order to obtain a  two-dimensional (2D) histogram.  In  Fig.~\ref{histogramma}, a 2D histogram of the band gap and layer thickness for sample~A is shown by the color plot. The band gap profile is clearly seen in the plot and decreases closer to the surface, as it is expected from the growth procedure.

From the SNMS sputtering time dependence of the GGI ratio we calculated the band profile according to  Eq.~\ref{eq_bandgap}. The best correspondence of the measured band gap with the profile calculated from composition of the sample~A is found with a bowing factor b~=~0.12 and is shown in Fig.~\ref{histogramma} by a continuous line. A good agreement of the measured profiles proves the accuracy of the new measurement technique and brings a precise layer thickness scale to the SNMS measurements.

Surface roughness of about 150~nm prevents interference measurements at the absorber layer thickness values above 2.2~$\mu m$  in the current sample. An increase in the band gap is not observed in this region due to the fact that CIGS surface defects effectively lower the band gap in the PL measurements near the surface of the unetched sample.

\begin{figure}[t]
\centerline{\includegraphics[width=3.375in]{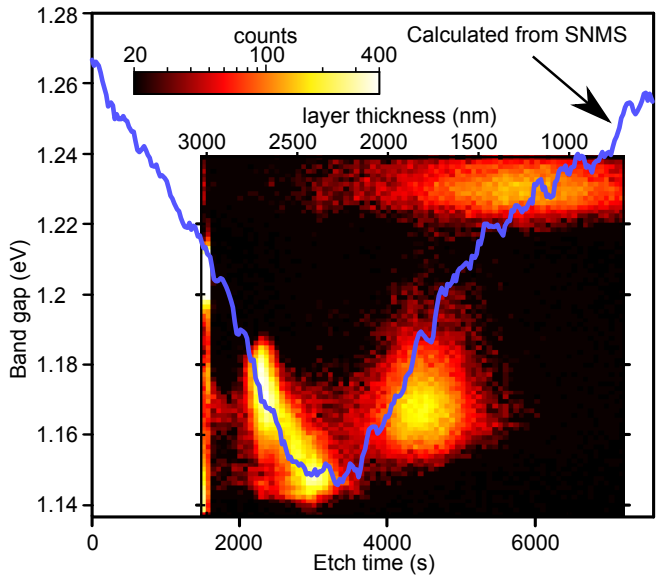}}
\caption{\label{histogrammb}Similar data representation as in Fig.~\ref{histogramma} but for the sample~B - 190,000 spectral measurements.}
\end{figure}

In Fig.~\ref{histogrammb} we compare the band gap and the layer thickness 2D histogram with the band gap calculated from the SNMS composition profile for the sample~B . Similar to the previous sample, a good correspondence between the band profiles is observed and we are able to extract quantitative information. The bowing factor for the sample~B is found to be b~=~0.215, significantly different than that for the sample~A. This demonstrates the importance of direct measurements. The surface roughness has a stronger effect on the interference measurements in the thick absorber layer of the sample~B.

To conclude, insights into the optoelectronic properties in the depth profile of thin-film solar cells are revealed using a new technique combining simultaneous measurements of a semiconductor film thickness and its band gap. An experimentally proven sample dependence of the bowing factor represents an important finding for reliable computation of the solar cells properties. A direct and accurate method fro measurements of the band gap grading in solar cells is presented, which has great potential for the use in other optoelectronic thin film devices.

The authors would like to thank to Theresa Magorian Friedlmeier for useful contribution, as well as the Concept for the Future in the Excellence Initiative at KIT for financial support. Z-H. Zhang acknowledges support by the Karlsruhe School of Optics \& Photonics (KSOP).

%% End of main article.



\end{document}